  \documentclass[final,5p,times,twocolumn]{elsarticle}

\usepackage{amssymb,amsmath}
\usepackage{verbatim}
\usepackage{color}
\usepackage{ulem}

\journal{Physics Letters B}

\begin{document}

\begin{frontmatter}



\title{Self-dual classical higher-spin multicopy}


\author[first]{Nikita Misuna}
\affiliation[first]{organization={I.E. Tamm Theory Department, Lebedev Physical Institute},
            addressline={Leninskiy prospekt 53}, 
            city={Moscow},
            postcode={119991}, 
            country={Russia}}
            
            \author[first,second]{Dmitry Ponomarev}
\affiliation[second]{organization={Institute for Theoretical and Mathematical Physics, Lomonosov Moscow State University},
            addressline={Leninskie Gory, GSP-1}, 
            city={Moscow},
            postcode={119991}, 
            country={Russia}}
            
                     \author[third]{Alexander Solomin}
\affiliation[third]{organization={Moscow Institute of Physics and Technology},
            addressline={Institutskiy pereulok 9}, 
            city={Dolgoprudny},
            postcode={141701}, 
            country={Russia}}

\begin{abstract}
We show that the  self-dual classical  double copy can be straightforwardly extended to the higher-spin case when formulated in terms of light-cone gauge prepotentials. This allows us to construct a higher-spin extension for any self-dual spacetime that admits {a} Kerr-Schild form.
 We also discuss the counterpart of this procedure at the level of Weyl tensors.  
  We find that{,} depending on the class of the original gravitational background{,} higher-spin Weyl tensors may follow various multicopy patterns.
\end{abstract}




\end{frontmatter}




\section{Introduction}
\label{introduction}
Massless higher-spin theories are notorious for difficulties with interactions \citep{Weinberg:1964ew,Coleman:1967ad}.  Nevertheless,  it was found not long ago \citep{Ponomarev:2016lrm} that local and interacting massless higher-spin theories  do exist in flat space, but these have to be self-dual \citep{Ponomarev:2017nrr}. This result relies on the earlier analysis \citep{Metsaev:1991mt,Metsaev:1991nb} of higher-spin interactions   carried out in the light-cone gauge formalism. It also parallels other earlier observations
that massless higher-spin fields can consistently propagate on self-dual gravitational and Yang-Mills backgrounds \citep{Aragone:1979hx,Eastwood:1981jy}
and that self-dual Yang-Mills theories may have an arbitrary number of supersymmetries \citep{Devchand:1996gv}.

In the present paper we focus on the analysis of classical solutions to chiral higher-spin theory equations of motion. To be more precise, we will deal with a two-derivative theory originally found in \citep{Ponomarev:2017nrr} as a result of a contraction -- which is analogous to the contraction that relates the Weyl-Moyal and the Poisson brackets -- of the original theory \citep{Ponomarev:2016lrm} and referred to as  the Poisson chiral higher-spin theory. A closely related theory was formulated in terms of a covariant action in \citep{Krasnov:2021nsq}, where it was referred to as higher-spin self-dual gravity.

Instead of solving the chiral higher-spin equations of motion directly, we will explore whether the classical double copy -- a duality that relates solutions of gravitational and gauge theories -- can be extended to higher spins. We will refer to such an extension as multicopy.
Once it is found, we will be able to generate classes of solutions to chiral higher-spin theory equations of motion by simply multicopying lower-spin solutions. 
One advantage of this approach is that higher-spin solutions found this way will automatically receive {an} interpretation as higher-spin extensions of respective lower-spin solutions.  Interpretation of higher-spin solutions directly, {e.g.,} by their geometric features,  is complicated by the absence of consistent higher-spin geometry \citep{Ammon:2011nk,Vasiliev:2014vwa,Ivanovskiy:2025kok,Ivanovskiy:2025ial}.

Double copy originates from string theory \citep{Kawai:1985xq} and it was initially formulated at the level of amplitudes \citep{Bern:2008qj,Bern:2010ue}. It was then carried over to classical solutions, where it can be formulated both in terms of gauge potentials \citep{Monteiro:2014cda} and in terms of Weyl tensors \citep{Luna:2018dpt}. Related earlier results can be found in \citep{Walker:1970un,Hughston:1972qf,Didenko:2008va}. For further details and references on the classical double copy we refer the reader to the book \citep{White:2024pve}.

Quite remarkably, the setup of self-dual theories formulated in the light-cone gauge not only enables the construction of local higher-spin theories, but  it is also special in the double-copy context, as it allows one to make the amplitude double copy manifest at the level of {the} action \citep{Monteiro:2011pc}. These results can be straightforwardly extended to higher spins \citep{Ponomarev:2017nrr,Monteiro:2022lwm,Ponomarev:2024jyg}. In addition, for self-dual theories one can define an alternative classical double copy \citep{Monteiro:2014cda,Berman:2018hwd,Luna:2018dpt}, which is related to the amplitude double copy in a more transparent manner. An observation that solutions in self-dual theories can be related in this way was first made in   \citep{Tod:1982mmp},
while some further developments of this approach can be found in \citep{Sabharwal:2019ngs,Elor:2020nqe,Chacon:2020fmr,Campiglia:2021srh,Armstrong-Williams:2022apo,Brown:2023zxm,Kim:2024dxo,Ilderton:2025gug}.

Extensions of the classical double copy to higher spins  have been explored before \citep{Didenko:2008va,Didenko:2011ir,Didenko:2021vui,Didenko:2022qxq,Brown:2025xlo}. These results deal with the standard classical double copy  and  are limited to free theories. Similarly, free massless higher-spin fields in the classical double-copy context were  discussed from the twistorial and  Newman-Penrose formalism
perspectives \citep{White:2020sfn,Chacon:2021wbr,Albertini:2025ogf}, and propagation of massless higher-spin fields on backgrounds related by  double copy was studied in \citep{Adamo:2023fbj}. Besides that, in \cite{Didenko:2021vdb}  certain solutions to Vasiliev’s equations  \citep{Vasiliev:1990en}  at the first non-trivial order in interactions were found in the Weyl multicopy form.
In this paper we will focus on the extensions of the self-dual classical double copy and their application to chiral higher-spin theories.

Solutions to higher-spin equations of motion have been studied previously within different frameworks. In particular, various classical solutions  -- including those exhibiting some multicopy features -- in Vasiliev's theory were found in \citep{Iazeolla:2007wt,Didenko:2009td,Iazeolla:2011cb,Didenko:2021vdb,Didenko:2025xca}.  
In three dimensions{,} matter-free higher-spin theories admit the Chern-Simons form \citep{Blencowe:1988gj} and the associated equations of motion can be solved by employing  standard methods
 \citep{Gutperle:2011kf,Ammon:2011nk,Castro:2011iw,Bunster:2014mua}.

The literature on classical solutions to chiral higher-spin theory equations of motion is more limited. Namely,   by adding the appropriate lower-spin corrections{,} the  BPST instanton was promoted to an exact solution of chiral higher-spin theory in \citep{Skvortsov:2024rng}. 
 In addition, a rich class of pp-wave solutions to chiral higher-spin theory equations of motion was found in \citep{Tran:2025yzd} and  the extension of the Taub-NUT solution to higher spins was constructed perturbatively in \citep{Skvortsov:2025ohi}. Finally, in \citep{Lang:2025rxt} a perturbative solution to the coloured chiral higher-spin theory equations of motion was found by a straightforward generalisation of the self-dual Yang-Mills construction of  \citep{Bardeen:1995gk,Rosly:1996vr}. All these results employ  covariant reformulations \citep{Krasnov:2021nsq,Sharapov:2022faa} of chiral higher-spin theories. In contrast to that, we will use the original light-cone form of the theory.

This letter is organised as follows. We start by reviewing the self-dual classical double copy in section \ref{review}. Next, in section \ref{hsprepotential} we present its extension to higher spins at the level of light-cone prepotentials. In section \ref{hsWeyl} the counterpart of this discussion at the level of Weyl tensors is given. We then conclude in section \ref{conclusions}.

\section{Self-dual classical double copy}
\label{review}

In this section we briefly review the  self-dual classical double copy for lower-spin fields. 
For further details, see \citep{Tod:1982mmp,Monteiro:2014cda,Berman:2018hwd,Luna:2018dpt}.
In what follows we will work with four-dimensional Minkowski space in light-cone coordinates
\begin{equation}
\label{24mar0}
ds^2 = 2 du dv - 2 dx dy.
\end{equation}

In  light-cone gauge{,} the equations of motion for self-dual Yang-Mills theory read \citep{Leznov:1986up,Parkes:1992rz,Chalmers:1996rq}
\begin{equation}
\label{2apr1}
\Box \Phi^{+1} + 2ig [\partial_v \Phi^{+1},\partial_x \Phi^{+1}]=0,
\end{equation}
where $\Phi^{+1}$ is a Lie algebra valued light-cone prepotential carrying helicity $+1$, $[\cdot, \cdot]$ is the Lie bracket and $g$ is the Yang-Mills coupling constant.
It is convenient to rewrite these as
\begin{equation}
\label{24mar1}
\Box \Phi^{+1} + \frac{g}{2}[D_{\dot\alpha}\Phi^{+1},D^{\dot\alpha}\Phi^{+1}]=0,
\end{equation}
where
\begin{equation}
\label{2apr2}
 D_{\dot\alpha} \equiv o^{\alpha}\partial_{\alpha\dot\alpha}, \quad \partial_{\alpha\dot\alpha}\equiv (\sigma^{\mu})_{\alpha\dot\alpha}\partial_\mu, \quad o^\alpha \equiv \delta^\alpha_1,
\end{equation}
$\sigma^\mu$ are the Pauli matrices and $o^\alpha$ is a fixed constant spinor.
The spinor indices are raised and lowered with the Levi-Civita tensor  $\varepsilon_{\dot\alpha\dot\beta}$\footnote{Our conventions for dealing with spinors are the same as in the amplitude literature, see, e.g., \citep{Elvang:2013cua}.}.
The Yang-Mills potential is connected to the light-cone prepotential via
\begin{equation}
\label{24mar3}
A_u = \partial_x \Phi^{+1}, \qquad A_y = \partial_v \Phi^{+1}, \qquad A_{ x} = A_v=0.
\end{equation}
For self-dual Maxwell theory one simply has
\begin{equation}
\label{24mar3x1}
\Box \Phi^{+1} =0.
\end{equation}

Similarly, the light-cone gauge version of self-dual gravity equations of motion has the form \citep{Plebanski:1975wn,Plebanski:1996np,Chalmers:1996rq}
\begin{equation}
\label{24mar4}
\Box \Phi^{+2} + \kappa D^{\dot\alpha}D^{\dot\alpha}\Phi^{+2} D_{\dot\alpha}D_{\dot\alpha}\Phi^{+2}=0,
\end{equation}
where $\Phi^{+2}$ is the gravitational light-cone prepotential carrying helicity $+2$ and $\kappa$ is the gravitational coupling constant. In (\ref{24mar4}) we used the standard convention, which implies that indices denoted by the same letter are symmetric. The fluctuation of the gravitational potential $h_{\mu\nu} \equiv g_{\mu\nu}-\eta_{\mu\nu}$ is related to prepotential $\Phi^{+2}$ via
\begin{equation}
\label{24mar5}
h_{uu} = -2 \partial^2_x \Phi^{+2}, \quad h_{uy} = -2 \partial_x\partial_v \Phi^{+2}, \quad 
h_{yy} = -2 \partial^2_v  \Phi^{+2}
\end{equation}
with other components vanishing. 

For metrics that admit Kerr-Schild form{,} one has 
\begin{equation}
\label{24mar6}
h_{\mu\nu} = k_\mu k_\nu \varphi
\end{equation}
for some vector field $k_\mu$ and a scalar field $\varphi$\footnote{Strictly speaking, {the} Kerr-Schild form implies that certain conditions  on $k$ and $\varphi$ are imposed. For this and other standard discussions on exact solutions in General Relativity we refer the reader to \citep{Stephani:2003tm}.}. As it is not hard to see, for the gravitational potential fluctuations as in (\ref{24mar5}){,}
this requirement translates into 
\begin{equation}
\label{24mar7}
D^{\dot\alpha}D^{\dot\alpha}\Phi^{+2} D_{\dot\alpha}D_{\dot\alpha}\Phi^{+2}=0.
\end{equation}
 By plugging (\ref{24mar7}) into (\ref{24mar4}) one finds
\begin{equation}
\label{24mar8}
\Box \Phi^{+2} =0.
\end{equation}
In other words, for self-dual Kerr-Schild metrics the two terms in (\ref{24mar4}) vanish separately and the gravitational prepotential is harmonic.

This observation allows one to relate any self-dual gravity background that admits Kerr-Schild form to a solution of {the} self-dual Maxwell equations (\ref{24mar3x1}) simply by stating that the associated light-cone prepotentials are equal
\begin{equation}
\label{24mar9}
\Phi^{+1}=c^{+1}\Phi^{+2},
\end{equation}
possibly up to an arbitrary constant factor $c^{+1}$. Equation (\ref{24mar9}) is the self-dual classical double copy relation with $\Phi^{+1}$ being the single copy of $\Phi^{+2}$. Obviously, in a similar manner one can define 
\begin{equation}
\label{24mar10}
\Phi^{0}=c^{0}\Phi^{+2},
\end{equation}
which solves the free massless scalar equation of motion and represents the associated zeroth copy. 

We briefly note that the standard Kerr-Schild double copy \citep{Didenko:2008va,Monteiro:2014cda}  for (\ref{24mar6}), instead of (\ref{24mar9}) and (\ref{24mar10}), gives the following single and zeroth copies
\begin{equation}
\label{2apr3}
\tilde A_\mu = k_\mu \varphi, \qquad  \tilde\Phi^0 = \varphi.
\end{equation}
In general, (\ref{24mar9}), (\ref{24mar10}) and (\ref{2apr3}) give different results \citep{Luna:2018dpt}. At the same time, the self-dual classical double copy can be interpreted as a version of the Kerr-Schild one in which one replaces the Kerr-Schild vector $k_\mu$ with a differential operator
\begin{equation}
\label{2apr4}
(\hat{k}_u,\hat{k}_v,\hat{k}_x,\hat{k}_y)= -2(\partial_x, 0,0,\partial_v),
\end{equation}
see (\ref{24mar3}), (\ref{24mar5}). As noted in \citep{Monteiro:2014cda}, the replacement of the Kerr-Schild vectors with differential operators makes the connection between the classical and the amplitude double copies more transparent. 

\subsection{Weyl double copy}
The self-dual classical double copy was also explored at the level of Weyl tensors \citep{Luna:2018dpt}. In this context the gauge theory field strength is regarded as the spin-1 counterpart of the gravitational Weyl tensor. Employing the background frame field $e_{\mu}{}^a$, 
\begin{equation}
\label{2apr5x00}
e_\mu{}^a e_\nu{}^b \eta_{ab}=g_{\mu\nu},
\end{equation}
 it can be converted to the local Lorentz basis 
\begin{equation}
\label{24mar11}
F_{ab} = (e^{-1})^\mu{}_a (e^{-1})^\nu{}_b F_{\mu\nu}.
\end{equation}
Then, by using the Pauli matrices{,} it can be expressed as a spin-tensor
\begin{equation}
\label{24mar12}
F_{ab}(\sigma^a)_{\alpha\dot\alpha}(\sigma^b)_{\beta\dot\beta} \equiv F_{\alpha\beta,\dot\alpha\dot\beta} =
\varepsilon_{\alpha\beta}\bar{C}_{\dot\alpha\dot\beta}+\varepsilon_{\dot\alpha\dot\beta}C_{\alpha\beta},
\end{equation}
where the last equality follows from $F_{ab}=-F_{{ba}}${;} moreover, both $C_{\alpha\beta}$ and $\bar{C}_{\dot\alpha\dot\beta}$ are symmetric and complex conjugate to each other for real fields in Lorentzian signature. 
In these terms self-duality amounts to $C_{\alpha\beta}=0$, which requires either complex fields or changing the signature to $(+,+,+,+)$ or $(+,+,-,-)$.

In a similar fashion one deals with gravity. In vacuum spacetimes the Riemann tensor $R_{\mu\nu\lambda\rho}$ is traceless and is equal to the Weyl tensor $R_{\mu\nu\lambda\rho} = W_{\mu\nu\lambda\rho}$. It is then contracted with the inverse frame fields
\begin{equation}
\label{24mar13}
W_{abcd} = (e^{-1})^\mu{}_a (e^{-1})^\nu{}_b(e^{-1})^\lambda{}_c (e^{-1})^\rho{}_d W_{\mu\nu\lambda\rho}.
\end{equation}
  Finally, when converted to spinors, the Weyl tensor decomposes as
 \begin{equation}
 \label{24mar14}
 W_{\alpha\beta\gamma\delta,\dot\alpha\dot\beta\dot\gamma\dot\delta} =\varepsilon_{\alpha\beta}
 \varepsilon_{\gamma\delta}\bar{C}_{\dot\alpha\dot\beta\dot\gamma\dot\delta}
 +\varepsilon_{\dot\alpha\dot\beta}
 \varepsilon_{\dot\gamma\dot\delta} C_{\alpha\beta\gamma\delta},
 \end{equation}
 where both $C_{\alpha\beta\gamma\delta} $ and $\bar{C}_{\dot\alpha\dot\beta\dot\gamma\dot\delta}$ are totally symmetric. 
In these terms self-duality amounts to $C_{\alpha\beta\gamma\delta}=0$. Below we will focus on self-dual Maxwell fields and self-dual spacetimes only. 
 
 The Weyl double copy relies on the factorisation of Weyl tensors\footnote{We will use ''Weyl tensors'' {irrespective} of whether the spinor representation is used.} into a symmetrised product of four rank-1 spinors
 \begin{equation}
 \label{24mar15}
 \bar{C}_{\dot\alpha\dot\alpha\dot\alpha\dot\alpha} = \mu_{\dot\alpha}\nu_{\dot\alpha}\tau_{\dot\alpha}\lambda_{\dot\alpha},
  \end{equation}
  which are referred to as principal spinors. 
 For Petrov type-D spacetimes  Weyl tensors are algebraically special in the sense that  in (\ref{24mar15}) two pairs of spinors are equal
  \begin{equation}
 \label{25mar1}
 \bar{C}_{\dot\alpha\dot\alpha\dot\alpha\dot\alpha} = \mu_{\dot\alpha}\mu_{\dot\alpha}\tau_{\dot\alpha}\tau_{\dot\alpha}.
  \end{equation}
 In this case, the standard Weyl double copy states that the Weyl tensor can be factorised in terms of {a} spin-1 field strength $\bar{C}_{\dot\alpha\dot\alpha}$ and {a} scalar field $C$ as 
  \begin{equation}
 \label{25mar1x1}
 \bar{C}_{\dot\alpha\dot\alpha\dot\alpha\dot\alpha} =\frac{1}{C} \bar{C}_{\dot\alpha\dot\alpha}\bar{C}_{\dot\alpha\dot\alpha},
  \end{equation}
 where both $\bar{C}_{\dot\alpha\dot\alpha}$ and $C$ give solutions to the equations of motion in a curved background defined by  Weyl tensor (\ref{25mar1}).
 Explicitly, one has 
  \begin{equation}
 \label{25mar2}
 \bar{C}_{\dot\alpha\dot\alpha} = 2^{-\frac{1}{6}}{\mu_{\dot\alpha} \tau_{\dot\alpha}}[\mu\tau]^{\frac{1}{3}}, \qquad C = 2^{-\frac{1}{3}} [\mu\tau]^{\frac{2}{3}},
 \end{equation}
  where $[\mu\tau] \equiv {\mu}_{\dot\alpha}{\tau}^{\dot\alpha}$.

In other words, {as far as the tensorial structure is concerned}, the standard Weyl double copy implies that the gravitational Weyl tensor equals the tensor square of  the Maxwell field strength. In contrast to that, 
as was found in \citep{Luna:2018dpt} for the Eguchi-Hanson instanton, which is a self-dual type-D spacetime, the self-dual single copy procedure as defined in (\ref{24mar9}) leads to {the} {Maxwell} tensor of the form
\begin{equation}
\label{25mar3}
 \bar{C}'_{\dot\alpha\dot\alpha} = \bar\mu_{\dot\alpha}\bar\mu_{\dot\alpha}.
\end{equation}
This suggests {replacing} (\ref{25mar1x1}) with 
  \begin{equation}
 \label{25mar4}
 \bar{C}_{\dot\alpha\dot\alpha\dot\alpha\dot\alpha} =\frac{1}{C'} \bar{C'}_{\dot\alpha\dot\alpha}\bar{C''}_{\dot\alpha\dot\alpha},
  \end{equation}
where $\bar{C'}_{\dot\alpha\dot\alpha}$ and $\bar{C''}_{\dot\alpha\dot\alpha}$ are associated with two different solutions of Maxwell's equations. 
Accordingly, the associated Weyl double copy is referred to as mixed.

\section{Self-dual multicopy for prepotentials}
\label{hsprepotential}

In this section we will explain how the self-dual classical double copy extends to higher-spin theories at the level of prepotentials. We will deal with a chiral higher-spin theory given by an action
\begin{equation}
\label{25mar5}
\begin{split}
S=& \int d^4x \sum_{h\in \mathbb{Z}} \Phi^{-h}\Box \Phi^h\\
+&2gl  \int d^4x \sum_{h_1,h_2,h_3\in \mathbb{Z}} \delta^{h_1+h_2+h_3}_2 \Phi^{h_1} D^{\dot\alpha}D^{\dot\alpha}\Phi^{h_2} D_{\dot\alpha}D_{\dot\alpha}\Phi^{h_3},
\end{split}
\end{equation}
which can be obtained by a straightforward contraction  \citep{Ponomarev:2017nrr} of the theory suggested originally in \citep{Ponomarev:2016lrm}. 
Action (\ref{25mar5}) features an infinite set of fields labelled by integer helicities $h$.
By varying the action one finds
\begin{equation}
\label{25mar6}
\Box \Phi^{h_1} + gl  \sum_{h_2+h_3=2-h_1}D^{\dot\alpha}D^{\dot\alpha}\Phi^{h_2} D_{\dot\alpha}D_{\dot\alpha}\Phi^{h_3}=0.
\end{equation}

The self-dual classical double copy (\ref{24mar9}), (\ref{24mar10}) suggests a natural higher-spin generalisation 
\begin{equation}
\label{25mar7}
\Phi^{h}=c^{h}\Phi^{+2}
\end{equation}
with $c^h$ being arbitrary constants. Then it is trivial to see that for $\Phi^{+2}$ satisfying {the} Kerr-Schild condition (\ref{24mar7}),
\begin{equation}
\label{25mar8}
D^{\dot\alpha}D^{\dot\alpha}\Phi^{h_2} D_{\dot\alpha}D_{\dot\alpha}\Phi^{h_3}=0
\end{equation}
holds for any $h_2$ and $h_3$. Similarly, all $\Phi^{h_1}$ are harmonic
\begin{equation}
\label{25mar9}
\Box \Phi^{h_1} =0
\end{equation}
as a consequence {of} (\ref{24mar8}). As a result, both terms in  (\ref{25mar6}) vanish separately and the self-dual multicopy (\ref{25mar7}) does 
give  solutions to {the} chiral higher-spin equations of motion (\ref{25mar6}).

 Putting it differently, via ansatz the (\ref{25mar7}) we {find} an infinite family of solutions to {the} chiral higher-spin theory equations of motion, which extends any self-dual gravitational background that admits {a} Kerr-Schild form to higher spins and features one independent integration constant $c^h$ for every helicity.

\section{Self-dual Weyl multicopy}
\label{hsWeyl}

In this section we will discuss to what extent double copy relations between Weyl tensors of different spins, such as (\ref{25mar4}), 
extend to higher spins, once the self-dual multicopy relation at the level of prepotentials (\ref{25mar7}) is adopted.

Before addressing this question{,} we should first clarify what we will mean by saying that the classical Weyl multicopy holds. 
One important feature of {the} lower-spin double copy (\ref{25mar1x1}), (\ref{25mar4}) is that Weyl tensors for different spins are related algebraically. Accordingly, if higher-spin Weyl tensors associated with prepotentials (\ref{25mar7}) can be expressed in terms of the components of the gravitational Weyl tensor algebraically{,} we may say that the higher-spin Weyl multicopy  holds.
Another distinguishing feature of the lower-spin Weyl double copy is that Weyl tensors admit a factorised form. Therefore, if higher-spin Weyl tensors factorise into Weyl tensors of lower-spin fields, we will say that Weyl multicopy holds in this second sense. 

As it is not hard to see, these two properties are, in principle, independent. For example, it may happen that the components of a higher-spin Weyl tensor can be expressed algebraically in terms of lower-spin ones, but the rank-1 spinors into which higher-spin Weyl tensors factorise are different from those appearing for the gravitational Weyl tensor. Vice versa, higher-spin Weyl tensors may involve the same set of principal spinors as the lower-spin ones, but also feature overall factors, which are algebraically independent of them. 
It turns out that, indeed, the aforementioned two properties can be  either satisfied or violated in different combinations depending on the class of the gravitational solution we start with in (\ref{25mar7}). 

Before moving  to higher spins{,} we recall that the self-dual component of Maxwell's field strength is related to the light-cone prepotential via, see{,} {e.g.,} \citep{Tod:1982mmp}
\begin{equation}
\label{26mar1}
\bar{C}_{\dot\alpha\dot\alpha}=D_{\dot\alpha}D_{\dot\alpha}\Phi^{+1}.
\end{equation}
At the same time, for self-dual gravity one has
\begin{equation}
\label{26mar2}
\bar{C}_{\dot\alpha\dot\alpha\dot\alpha\dot\alpha}=D_{\dot\alpha}D_{\dot\alpha}D_{\dot\alpha}D_{\dot\alpha}\Phi^{+2}.
\end{equation}
We would like to emphasise that this formula, despite being linear in $\Phi^{+2}$, is valid not only for linearised gravity, but also holds for general self-dual spacetimes \citep{Plebanski:1975wn}.

Higher-spin curvatures were originally defined in \citep{deWit:1979sib} for free theories in terms of Fronsdal fields \citep{Fronsdal:1978rb}.
Within the frame formalism higher-spin curvatures and the associated Weyl tensors were introduced in \citep{Vasiliev:1986zej}.
For our purposes it is more convenient to use \citep{Penrose:1965am}
\begin{equation}
\label{26mar3}
\bar{C}_{\dot\alpha(2s)}=(D_{\dot\alpha})^{2s}\Phi^{+s},
\end{equation}
which immediately connects higher-spin Weyl tensors with the associated light-cone prepotentials for positive-helicity fields. 
In (\ref{26mar3}) it is understood that the higher-spin Weyl tensor has $2s$ spinorial indices and it is totally symmetric.
An analogous formula is valid for negative-helicity fields. In what follows, we will focus on positive-helicity fields only.

To proceed, we need a non-linear version of (\ref{26mar3}), which would be valid for higher-spin theory (\ref{25mar5}).
Unfortunately, it is not readily available in the literature. Higher-spin Weyl tensors for the chiral higher-spin theory were introduced in \citep{Sharapov:2022faa} in the process of covariantisation of the theory. To obtain the desired formula one would need to carry out the Poisson contraction of the covariant equations of \citep{Sharapov:2022faa}  and then perform the light-cone gauge fixing. 
We leave {the} rigorous discussion of this issue for future work, while in the present paper we will keep (\ref{26mar3}) as an assumption. 

With all these preliminaries settled{,} we can finally reduce the question of the existence of the self-dual  Weyl multicopy to a formal mathematical problem. Namely, considering that the higher-spin prepotentials are proportional to the gravitational one (\ref{25mar7}) 
and that the higher-spin Weyl tensors are given by higher derivatives of these prepotentials (\ref{26mar3}) in the $(x,v)$ plane, the problem of the existence of the Weyl multicopy amounts to the question whether higher derivatives of $\Phi^{+2}$ are expressible in terms of the first four derivatives -- which are, in turn,  associated with the gravitational Weyl tensor -- provided {the} Kerr-Schild condition (\ref{25mar8}) is satisfied.

This problem can be efficiently addressed by applying the machinery of unfolding \citep{Vasiliev:1988xc} to the Kerr-Schild condition (\ref{25mar8}) regarded as a differential equation in two-dimensional space.
The resulting analysis is somewhat technical and requires a  laborious case{-}by{-}case study. It will be presented in a subsequent paper. 
Below we will make some general remarks, give a qualitative overview of the procedure and briefly present its results. It should be mentioned that the first steps of this procedure were already carried out in \citep{Tod:1982mmp}.

\subsection{Unfolding {the} Kerr-Schild condition}

Even without resorting to any special techniques{,} one can easily see that {the} Kerr-Schild condition is not sufficient to express higher derivatives of $\Phi^{+2}$ in terms of a finite number of its lower derivatives. Indeed,  the Kerr-Schild condition itself puts a single constraint on the three second derivatives $D_{\dot\alpha}D_{\dot\alpha}\Phi^{+2}$, which means that two of them are unconstrained. Similarly, by differentiating {the} Kerr-Schild condition $n$ times{,} one finds $n+1$ equations, which involve  derivatives of $\Phi^{+2}$ up to order $n+2$. In total, there are
$n+3$ derivatives of $\Phi^{+2}$ of order $n+2$ involved, which means that, in general, higher-derivative consequences of {the} Kerr-Schild condition leave two of them algebraically undetermined. 
This implies that {the} Weyl multicopy in the aforementioned sense for general Kerr-Schild backgrounds does not hold. 

This discussion can also be summarised by saying that a general solution of {the} Kerr-Schild condition (\ref{24mar7}){,} regarded as a differential equation in two variables{,} is determined by two functions of a single variable. In addition to that, we have to remember that 
$\Phi^{+2}$ is, in fact, a function of four variables, which also satisfies (\ref{24mar8}). The latter equation can always be solved for the $u$-dependence no matter what {the} $y$-dependence is. We, therefore, conclude that, in general,  self-dual Kerr-Schild backgrounds can be classified by two independent functions of two variables. This superficial counting agrees with the analysis of \citep{Tod:1982mmp}.

To analyse the problem systematically, we solve (\ref{24mar7}) as
\begin{equation}
\label{26mar4}
D_{\dot\alpha}D_{\dot\alpha}\Phi^{+2} = \mu_{\dot\alpha}\mu_{\dot\alpha}.
\end{equation}
Indeed, (\ref{24mar7}) just says that the determinant of a symmetric matrix $D_{\dot\alpha}D_{\dot\alpha}\Phi^{+2}$ vanishes.  
In order to rephrase {the} Kerr-Schild condition in terms of $\mu$, we need to take into account that $\mu_{\dot\alpha}\mu_{\dot\alpha}$ carries second derivatives of $\Phi^{+2}$. Since derivatives commute{,} (\ref{26mar4}) entails
\begin{equation}
\label{26mar5}
D^{\dot\alpha}(\mu_{\dot\alpha}\mu_{\dot\alpha})=0.
\end{equation}
By solving (\ref{26mar5}), we find 
\begin{equation}
\label{26mar6}
D_{\dot\alpha}\mu_{\dot\beta} = \lambda_{(\dot\alpha}\mu_{\dot\beta)}+\frac{1}{6}\varepsilon_{\dot\alpha\dot\beta}[\lambda\mu],
\end{equation}
where $\lambda$ is an arbitrary spinor {carrying} the third derivatives of $\Phi^{+2}$, that are algebraically independent of $\mu$.
Let us note that  (\ref{26mar6}) implies 
\begin{equation}
\label{2apr5x0}
D_{(\dot\alpha}\mu_{\dot\beta)}= \lambda_{(\dot\alpha}\mu_{\dot\beta)},
\end{equation}
therefore,
\begin{equation}
\label{2apr5}
(D_{\dot\alpha})^3\Phi^{+2} = 2\mu_{\dot\alpha}\mu_{\dot\alpha}\lambda_{\dot\alpha}.
\end{equation}
Differentiating (\ref{2apr5}) further, {irrespective} of whether {the} derivatives act on $\mu$ or on the remaining part of the expression, by (\ref{2apr5x0}) the factor of $\mu_{\dot\alpha}\mu_{\dot\alpha}$ is {always} present. Thus,  multicopied higher-spin Weyl tensors (\ref{26mar3}) always feature a factor of $\mu_{\dot\alpha}\mu_{\dot\alpha}$.

Next, we study  differential consequences of (\ref{26mar6}). To be more precise,  as a consequence of $D^{\dot\alpha}D_{\dot\alpha}\mu_{\dot\beta} =0${,} from (\ref{26mar6}) one finds a constraint on the derivatives of $\lambda$
\begin{equation}
\label{26mar7}
\mu^{\dot\alpha}D_{\dot\alpha}\lambda_{\dot\beta}=\frac{2}{3}[\lambda\mu]\lambda_{\dot\beta} - \frac{1}{2}D_{\dot\alpha}\lambda^{\dot\alpha}\mu_{\dot\beta}.
\end{equation}
Assuming that $[\lambda\mu]\ne 0$, the derivatives of $\lambda$ can be decomposed into a basis formed by symmetric products of $\lambda_{\dot\alpha}$ and $\mu_{\dot\beta}$ and by $\varepsilon_{\dot\alpha\dot\beta}$
\begin{equation}
\label{3apr1}
D_{\dot\alpha}\lambda_{\dot\beta}=\frac{2}{3}\lambda_{\dot\alpha}\lambda_{\dot\beta} - 2X \lambda_{(\dot\alpha}\mu_{\dot\beta)}+AX^2\mu_{\dot\alpha}\mu_{\dot\beta}-\frac{1}{2}\varepsilon_{\dot\alpha\dot\beta}\Lambda,
\end{equation}
where 
\begin{equation}
\label{3apr2}
\Lambda \equiv D_{\dot\alpha} \lambda^{\dot\alpha}, \qquad X\equiv \frac{\Lambda}{[\lambda\mu]}
\end{equation}
and $A$ is {a} new independent function. In other words, we found that, in general, out of {the} four first derivatives of $\lambda_{\dot\alpha}$ only two are algebraically independent. Above we chose them as $A$ and $\Lambda$.

As it is not hard to see, already here we encounter a special case $[\lambda\mu]=0$ for which decomposition (\ref{3apr1}) is not valid and the above analysis should be revisited. Namely,  $[\lambda\mu]=0$ implies $\lambda = \kappa_0 \mu$ for some function $\kappa_0${,}
and (\ref{26mar6}) acquires the form
\begin{equation}
\label{3apr3}
D_{\dot\alpha}\mu_{\dot\beta} = \kappa_0\mu_{\dot\alpha}\mu_{\dot\beta}.
\end{equation}
Then, $D^{\dot\alpha}D_{\dot\alpha}\mu_{\dot\beta} =0$ entails 
\begin{equation}
\label{3apr4}
D_{\dot\alpha}\kappa_0 = \mu_{\dot\alpha}\kappa_1,
\end{equation}
where $\kappa_1$ is an algebraically independent derivative of $\kappa_0$. The same pattern repeats at higher orders
\begin{equation}
\label{3apr5}
D_{\dot\alpha}\kappa_n = \mu_{\dot\alpha}\kappa_{n+1}, \quad n>0.
\end{equation}
In this case{,} for the gravitational Weyl tensor one has
\begin{equation}
\label{3apr6}
\bar{C}_{\dot\alpha(4)} = 2(\kappa_1+3\kappa_0^2)(\mu_{\dot\alpha})^4,
\end{equation}
so it is {of type N}. Differentiating it further, we find that by  (\ref{3apr5})  higher-spin Weyl tensors are of the form
\begin{equation}
\label{26mar8}
\bar{C}_{\dot\alpha(2s)}\propto (\mu_{\dot\alpha})^{2s},
\end{equation}
where at every order the proportionality coefficient involves a new algebraically independent function $\kappa_{2s-3}$.

Proceeding with (\ref{3apr1}) to higher orders{,} one finds further special branches of solutions to the unfolded equations for {the} Kerr-Schild constraint (\ref{24mar7}).
Our results can be briefly summarised as follows.

\subsection{Summary of results}

As we mentioned previously, for general self-dual Kerr-Schild spacetimes higher derivatives of the gravitational prepotential $\Phi^{+2}$ 
involve algebraically independent functions{;} thus, the self-dual Weyl multicopy does not hold. The only pattern that holds to all orders is that higher-spin Weyl tensors involve $\mu_{\dot\alpha}\mu_{\dot\alpha}$ as a factor. 

For general self-dual  type N  backgrounds that admit {a} Kerr-Schild form{,} the higher-spin Weyl tensors follow the pattern (\ref{26mar8}).
Thus, these factorise into the same principal spinors that enter the gravitational Weyl tensor decomposition and in this sense the multicopy holds. At the same time, the overall factor is not algebraically expressible in terms of the components of the  gravitational Weyl tensor.
Let us also remark that in this case the self-dual multicopy solves the equations of motion of the chiral higher-spin theory \citep{Ponomarev:2016lrm} with all higher-derivative interactions included. Indeed, at the level of the equations of motion, these are of the form
\begin{equation}
\label{9apr1}
(D^{\dot\alpha})^n\Phi^{h_2} (D_{\dot\alpha})^n\Phi^{h_3}
\end{equation}
and vanish as a consequence of (\ref{26mar3}) and (\ref{26mar8}).

For general self-dual  type D  backgrounds that admit {a} Kerr-Schild form{,} multicopied higher-spin Weyl tensors are
\begin{equation}
\label{26mar9}
\bar{C}_{\dot\alpha(2s)} =\mu_{\dot\alpha}\mu_{\dot\alpha}  \tau_{\dot\alpha}\tau_{\dot\alpha} F_{\dot\alpha(2s-4)}(\mu_{\dot\alpha},\tau_{\dot\alpha}),
\end{equation}
where $F$ is a fixed function. In other words, higher-spin Weyl tensors universally feature a factor of $\mu_{\dot\alpha}\mu_{\dot\alpha}  \tau_{\dot\alpha}\tau_{\dot\alpha}$, which is the gravitational Weyl tensor. Moreover, all components of higher-spin Weyl tensors are algebraically expressible in terms of spinors $\mu$ and $\tau$. At the same time, higher-spin Weyl tensors do not factorise into {a} product of $\mu$ and $\tau$. Instead, $F_{\dot\alpha(2s-4)}$ can be presented as a symmetric product of rank-1 spinors, which are certain linear combinations of $\mu$ and $\tau$.

Besides that, there are some further special cases. In particular, for the type-D case one can additionally  consistently impose
\begin{equation}
\label{26mar10}
D_{\dot\alpha}\lambda_{\dot\beta}=\frac{2}{3}\lambda_{\dot\alpha}\lambda_{\dot\beta}.
\end{equation}
This implies that fourth derivatives of $\Phi^{+2}$ carried by $D_{\dot\alpha}\lambda_{\dot\beta}$ are entirely expressed in terms of lower derivatives. As a consequence, all higher derivatives of $\Phi^{+2}$ are expressible in terms of its derivatives up to order three. As a result, one finds a closed formula for higher-spin Weyl tensors
\begin{equation}
\label{26mar11}
\bar{C}_{\dot\alpha(2s)}=\frac{2}{9} \left(\frac{2}{3} \right)^{2s-4} (2s)! \mu_{\dot\alpha}\mu_{\dot\alpha}(\lambda_{\dot\alpha})^{2s-2},
\end{equation}
which are not only algebraically expressible in terms of, but also factorise into the principal spinors $\mu$ and $\lambda$ that feature in the gravitational Weyl tensor. This case, in particular, covers the Eguchi-Hanson instanton \citep{Eguchi:1978xp}. A similar pattern holds for special type-N backgrounds, which include the Sparling-Tod instanton \citep{Sparling:1981nk}.

Finally, we note that in the spirit of the differential Kerr-Schild double copy (\ref{2apr4}), one may consider replacing the principal spinors appearing in the Weyl tensor decomposition (\ref{24mar15}) with differential operators. In this sense (\ref{26mar2}) tells us that any gravitational self-dual Weyl tensor is of differential Petrov type N. The associated multicopied higher-spin Weyl tensors (\ref{26mar3}) are then of the generalised differential type N. {Thus, we} find that for self-dual theories differential multicopy works far more naturally
not only at the level of potentials, but also at the level of Weyl tensors.

\section{Summary and conclusions}
\label{conclusions}

In the present paper we explored extensions of the self-dual classical double copy to chiral higher-spin theories. We found that it, indeed, straightforwardly extends to higher spins when formulated in terms of light-cone gauge prepotentials. To be more precise, for any 
self-dual background that admits {a} Kerr-Schild form the multicopy procedure amounts to equating -- possibly up to a constant factor -- higher-spin light-cone prepotentials to the gravitational prepotential associated with the original gravitational background. 
This allows us to find a large family of solutions to {the} chiral higher-spin theory equations of motion extending all these gravitational backgrounds to higher spins. 

We also explored this multicopy at the level of higher-spin Weyl tensors. These are expected to be given by higher derivatives of the higher-spin prepotentials, so the existence of the self-dual Weyl multicopy amounts to the analysis of whether higher derivatives of the gravitational prepotential can be expressed in terms of the first four derivatives carried by the gravitational Weyl tensor. We addressed this problem by using a version of the unfolding procedure and found that, in general, no multicopy pattern at the level of higher-spin Weyl tensors exists. 
At the same time, for specific backgrounds, depending on their class, higher-spin Weyl tensors do display some familiar Weyl double copy features. Our analysis also shows that the self-dual double copy does not entail any simple relations even between Weyl tensors of lower spins, in contrast to what the example of the Eguchi-Hanson instanton \citep{Luna:2018dpt} suggests.

 Let us also note some similarities between our analysis and the multipole expansion \citep{Geroch:1970cc}, which was recently discussed in the classical Weyl double-copy context in \citep{Chacon:2021hfe}.
Besides that,  the higher-spin solutions we found for {type-N} backgrounds seem to be closely related to the results  of \citep{Tran:2025yzd}. It would be interesting to explore this connection. 

One important loose end of our work is formula (\ref{26mar3}), which relates  Weyl tensors and light-cone prepotentials in chiral higher-spin theories. It was conjectured based on the expectation that, similarly to self-dual gravity, this relation is not affected by interactions. We hope to clarify this issue in {the} future.

The fact that, via the classical multicopy, we effortlessly managed to extend a very rich class of self-dual Kerr-Schild gravitational backgrounds to higher spins demonstrates {the} high efficiency of the multicopy as a solution-generating technique for {the} chiral higher-spin theory equations of motion and suggests numerous natural extensions. In particular, it would be interesting to explore whether the standard Kerr-Schild and Weyl double copies can be extended to chiral higher-spin theories. 
It would also be interesting to study how the classical double copy extends to the original chiral higher-spin theory  \citep{Ponomarev:2016lrm}, not to its lower-derivative contraction studied here. Similarly, one can explore relations between solutions of self-dual Yang-Mills theory and solutions of chiral higher-spin theories with internal symmetry.  A straightforward extension of our construction allows one to promote any solution of self-dual Yang-Mills equations that simultaneously solves Maxwell's equations to higher spins. It would be interesting to explore this issue more thoroughly.

Finally, as the extensive literature on the classical double copy suggests, it acts at many different levels \citep{Sabharwal:2019ngs,Alfonsi:2020lub,White:2020sfn,Campiglia:2021srh,Alawadhi:2021uie,Godazgar:2021iae,Chawla:2023bsu,Easson:2023dbk,Ferrero:2024eva} including, {e.g.,} isometries, horizons, topology, asymptotic symmetries, twistor geometry etc. At the same time, {the} geometry of classical higher-spin backgrounds {has remained} elusive so far. 
In this regard, one may expect that multicopy can be used to carry  the existing geometric constructions over from lower to higher spins, thus, facilitating progress in this direction.

\appendix



\biboptions{sort&compress}
\bibliographystyle{elsarticle-num} 
\bibliography{mult}






\end{document}